\documentclass[runningheads]{llncs}

\let\subparagraph\paragraph 
\usepackage{booktabs}
\usepackage{graphicx}
\usepackage{multirow}
\usepackage{caption}
\usepackage{subcaption}
\usepackage{array}

\usepackage{booktabs,collcell,eqparbox,makecell}
\usepackage{titlesec}

\def\Plus{\texttt{+}}

\titlespacing*{\section}{0pt}{1.1\baselineskip}{\baselineskip}

\begin{document}
\title{ConnectedUNets\Plus\Plus: Mass Segmentation from Whole Mammographic Images}
%
%

\author{{Prithul Sarker$^*$} \and
{Sushmita Sarker$^*$}\and
George Bebis \and
Alireza Tavakkoli}
\authorrunning{Sarker et al.}
%
\institute{Department of Computer Science and Engineering,\\ University of Nevada, Reno, United States \\
\email{\{prithulsarker, sushmita\}@nevada.unr.edu}}
%
\maketitle   
\def\thefootnote{*}\footnotetext{Prithul Sarker and Sushmita Sarker have equal contribution and are co-first authors}
%
\begin{abstract}
Deep learning has made a breakthrough in medical image segmentation in recent years due to its ability to extract high-level features without the need for prior knowledge. In this context, U-Net is one of the most advanced medical image segmentation models, with promising results in mammography. Despite its excellent overall performance in segmenting multimodal medical images, the traditional U-Net structure appears to be inadequate in various ways. There are certain U-Net design modifications, such as MultiResUNet, Connected-UNets and AU-Net, that have improved overall performance in areas where the conventional U-Net architecture appears to be deficient. Following the success of UNet and its variants, we have presented two enhanced versions of the Connected-UNets architecture: ConnectedUNets\Plus \space and ConnectedUNets\Plus\Plus. In ConnectedUNets\Plus, we have replaced the simple skip connections of Connected-UNets architecture with residual skip connections, while in ConnectedUNets\Plus\Plus, we have modified the encoder decoder structure along with employing residual skip connections. We have evaluated our proposed architectures on two publicly available datasets, the Curated Breast Imaging Subset of Digital Database for Screening Mammography (CBIS-DDSM) and INbreast. 

\keywords{Convolutional Neural Network \and Mammogram \and Semantic Segmentation \and U-Net \and ConnectedU-Nets \and MultiResUNet}
\end{abstract}
\section{Introduction}
Breast cancer is the most frequent type of cancer that causes death in women, with 44,130 instances reported in the United States in 2021~\cite{american2019breast}. The need for frequent mammography screening has been stressed in many studies in order to reduce mortality rates by finding breast malignancies before they spread to other normal tissues and healthy organs. 
A mammogram is an X-ray image of the breast to record changes in the tissue. 
The disease is typically identified by the presence of abnormal masses and microcalcifications in mammograms~\cite{elter2009cadx},~\cite{jiang1999improving}. Radiologists examine a high number of mammograms on a daily basis looking for abnormal lesions and assessing the location, shape, and type of any suspicious area in the breast. This is an important procedure which requires high precision and accuracy, however, it is still costly and prone to errors since detecting these regions is challenging as their pixel intensities often coincide with normal tissue. 


Deep learning advances~\cite{lecun2015deep}, especially Convolutional Neural Networks (CNN) \cite{zaheer2018gpu}, have shown a lot of promise in addressing these issues.
Despite being a game-changer in computer vision, CNN architectures have a key drawback: they require an enormous amount of training data. 
In order to solve this problem, U-Net~\cite{ronneberger2015u} is introduced which is built on a simple encoder-decoder network with multiple sets of CNN.
Even with a limited quantity of labeled training data, U-Net has demonstrated tremendous promise in segmenting breast masses, to the point where it has become the de-facto standard in medical image segmentation~\cite{litjens2017survey}. 
In light of the success of U-Net, various U-Net versions, such as Connected-UNets~\cite{baccouche2021connected} and AU-Net~\cite{sun2020aunet}, have been proposed. These variations have demonstrated promising results but appear to be inefficient in terms of fully recovering the region of interest in a given image. 

In this work, we have proposed and experimented with two enhanced versions of the Connected-UNets architecture. Although the proposed networks share an architectural similarity, they are designed for different use cases which are crucial in real-world scenario.
The proposed architectures take the entire mammogram image as input and perform mass segmentation along with mass boundary extraction.
The main contributions of our work include:

\begin{enumerate}
    \item We have proposed ConnectedUNets\Plus \space and ConnectedUNets\Plus\Plus, two novel and improved versions of the Connected-UNets, by utilizing residual skip connections and enhanced encoder-decoder in order to achieve better convergence.
    

    \item We have assessed the proposed architectures using full mammogram images in contrast to the baseline model which operates on cropped images of correctly detected and classified masses by an object detection model.

    \item We have experimented using all the images from two publicly available datasets, the Curated Breast Imaging Subset of Digital Database for Screening Mammography (CBIS-DDSM) ~\cite{lee2017curated} and INbreast ~\cite{moreira2012inbreast} for segmenting the region of interest (ROI) of breast mass tumors.
    
\end{enumerate}

To the best of our knowledge, our paper is the first to address the shortcomings of other papers' methodologies and to conduct an unbiased comparison.
We applied the same loss function, optimizer, and image size to all architectures to maintain objectivity. Additionally, to ensure a fair and accurate comparison, we used full mammograms as input for all the models and adopted a comparable preprocessing approach. 

\section{Related Works}
U-Net~\cite{ronneberger2015u}, a deep learning network having an encoder-decoder architecture, is among the most prominent deep neural networks commonly employed in medical image segmentation. 
The network has a symmetric architecture, with an encoder which extracts spatial information from the image and a decoder which constructs the segmentation map from the encoded data. 
The encoder and decoder are linked by a series of skip connections which are the most innovative component of the U-Net architecture since they enable the network to recover spatial data that has been lost due to pooling procedures. Abdelhafiz et al.~\cite{abdelhafiz2020convolutional} used a vanilla U-Net model to segment mass lesions in whole mammograms.
To segment suspicious regions in mammograms, Ravitha Rajalakshmi et al.~\cite{ravitha2021deeply} presented a deeply supervised U-Net model (DS U-Net) combined with a dense Conditional Random Field (CRF). 
Li et al.~\cite{li2018improved} proposed a Conditional Residual U-Net, named CRUNet, to improve the performance of the basic U-Net for breast mass segmentation.

Though U-Net is among the most popular and successful deep learning models for biomedical image segmentation, several improvements are still possible. 
Specifically, the concatenation of encoder and decoder features reveals a significant semantic gap despite the preservation of dispersed spatial features, which is a shortcoming of the simple skip connections. 
To deal with this issue, Ibtehaz et al.~\cite{ibtehaz2020multiresunet} proposed the MultiResUNet architecture by incorporating some convolutional layers along with shortcut connections in U-Net. Instead of simply concatenating the feature maps from the encoder stage to the decoder stage, they first pass them through a chain of convolutional layers and then concatenate them with the decoder features, which makes learning substantially easier.
This idea is inspired from the image-to-image conversion using convolutional neural networks~\cite{he2016deep}, where pooling layers are not favorable for the loss of information. MultiResUNet has shown excellent results on different biomedical images, however, the authors did not experiment with mammograms.  

Based on the U-Net architecture, Baccouche et al.~\cite{baccouche2021connected} proposed an improved architecture that connects two simple U-Nets, called Connected-UNets. In addition to the original idea of the U-Net architecture, which includes skip connections between the encoder and decoder networks, it cascades a second U-Net and adds skip connections between the decoder of the first U-Net and the encoder of the second U-Net. 
The key idea was to recovering fine-grained characteristics lost in U-Net's encoding process. 
However, the authors first used YOLO~\cite{redmon2016you} to detect the location of masses in mammograms, and then applied their method to segment only correctly localized masses. Such an approach is not optimum in practical settings where it is desirable to simultaneously localize and segment masses in whole mammograms rather than processing cropped mammograms. 

Several modifications of the U-Net architecture have also been proposed by incorporating an attention mechanism, which has shown to be extremely effective in medical image segmentation. Oktay et al.~\cite{oktay2018attention} proposed a new attention U-Net by adding an attention gate into the conventional U-Net. This enhanced the accuracy of the predictions. However, they didn't evaluate their model for breast mass segmentation. Similarly, Li et al.~\cite{li2019attention} built an attention dense U-Net for breast mass segmentation, which was compared to U-Net~\cite{ronneberger2015u}, Attention U-Net~\cite{oktay2018attention}, and DenseNet~\cite{hai2019fully}.
In another study by Sun et al.~\cite{sun2020aunet}, an attention-guided dense upsampling network, called AUNet, was built for breast mass segmentation in full mammograms. 
The major drawback of the papers mentioned above is they did not use all of the images available in the CBIS-DDSM dataset for experimention (i.e., they only used a portion of the images in the training and test sets). As a result, higher scores were reported in their studies. 

In this paper, we propose ConnectedUNets\Plus \space and ConnectedUNets\Plus\Plus, two enhanced versions of the Connected-UNets and Connected-ResUNets architectures by focusing on the limitations of the aforementioned architectures. An important contribution of our work is that we compared the proposed architectures with previously reported works under identical conditions. Moreover, we did not employ any object detection models for mass localization; instead, we conducted all of our experiments using whole mammograms.



\section{Methodology}

\subsection{Architecture}

\begin{figure}[b!]
\centering
\begin{subfigure}[b]{0.9\textwidth}
\centering
\includegraphics[width=\textwidth]{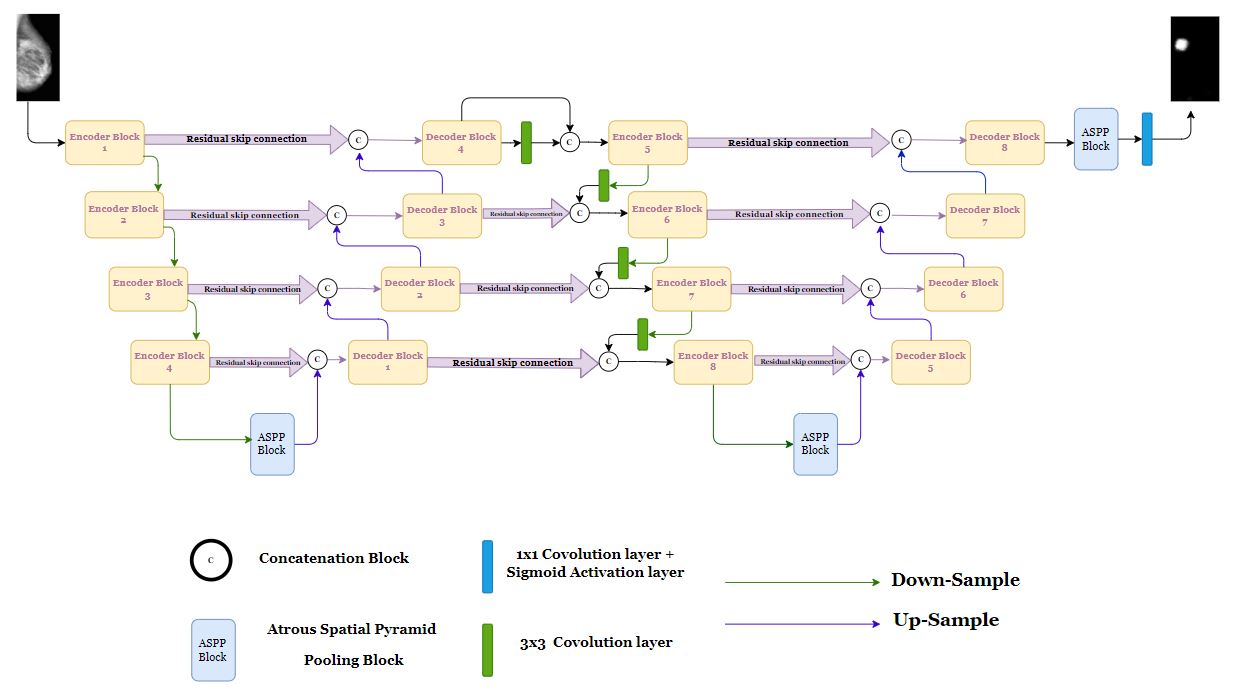}
\caption{Proposed architecture of ConnectedUNets\Plus\Plus}
\label{fig:(a)}
\end{subfigure}
\vspace{.6cm}

\begin{subfigure}[b]{0.9\textwidth}
\centering
\includegraphics[width=.9\textwidth]{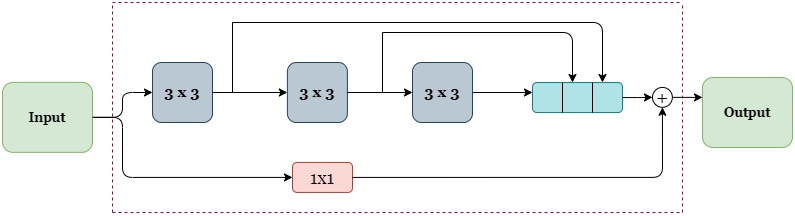}
\caption{Architecture of encoder and decoder blocks} \label{fig:(b)}
\end{subfigure}
\vspace{.6cm}

\begin{subfigure}[b]{0.9\textwidth}
\centering
\includegraphics[width=.9\textwidth]{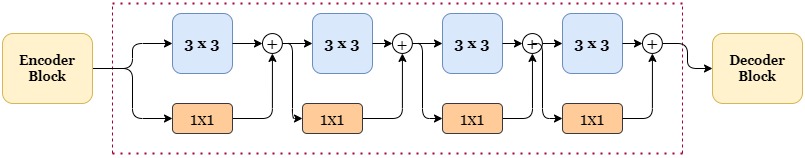}
\caption{Architecture of residual skip connections} \label{fig:(c)}
\end{subfigure}

\caption{Detailed ConnectedUNets\Plus\Plus Architecture}
\label{fig:Fig 1}
\end{figure}

We have used Connected-UNets~\cite{baccouche2021connected} as our baseline model because of its architectural elegance and performance. Even though, at first glance, ConnectedUNets\Plus \space and ConnectedUNets\Plus\Plus \space may merely seem a logical extension of Connected-UNets, the introduction of residual skip connections between the encoder and decoder is essential for successful segmentation. This not only improves the metric scores but also bridges the semantic barrier between the encoder-decoder features. The most crucial distinction is that ConnectedUNets\Plus \space and ConnectedUNets\Plus\Plus \space have been designed to enable mass segmentation from full mammograms rather than cropped mammograms. Figure~\ref{fig:(a)} shows an overview of our proposed ConnectedUNets\Plus\Plus \space architecture. Please take note that we have not included a separate illustration for the ConnectedUNets\Plus\space since both the architectures are identical with the exception of the encoder-decoder block, which has been maintained standard like the baseline model. For both the models, we have replaced the simple skip connections of Connected-UNets with more optimal residual skip connections between encoder and decoder and between the UNets as well. However, for the ConnectedUNets\Plus\Plus \space architecture, we have also modified the encoder-decoder block by including three 3x3 convolutions and one residual connection followed by an activation layer ReLU (Rectified Linear Unit) and a batch normalization (BN) layer as shown in Figure~\ref{fig:(b)}.
Specifically, the residual skip connections consist of four 3x3 convolutions where each of them is accompanied by one 1x1 convolution. The architecture of the residual skip connection is shown in Figure~\ref{fig:(c)}. The number of convolution blocks decreases in the deeper layer of the network as the semantic gap between encoder and decoder decreases due to getting closer to the bottleneck. Table~\ref{table-residual} shows the number of filters used in each of the residual skip connections for different layers.

\begin{table}[!ht]
  \centering
  \caption{ConnectedUNets\Plus\Plus \space and ConnectedUNets\Plus \space architecture details in terms of residual skip connections. These specifics apply to both architectures.\centering}
  \renewcommand{\arraystretch}{0.4}
  \scalebox{1}
  {
  \label{table-residual}
    \begin{tabular}{p{3.9cm}p{2cm}p{2.5cm}c}
    \toprule
    {\textbf{Residual Skip Connection}}\centering & {\textbf{No. of Conv. layer}}\centering & {\textbf{Conv. Layer Kernel Size}}\centering & {\textbf{No of Filters}} 
    \\
    \midrule
    {Residual Skip Connection}\centering & 4\centering & 3x3\centering  & 32 
    \\
    {01/08}\centering & 4\centering & 1x1\centering  & 32
    \\
    \midrule
    {Residual Skip Connection}\centering & 3\centering & 3x3\centering  & 64 
    \\
    {02/05/09}\centering & 3\centering & 1x1\centering  & 64 
    \\
    \midrule
    {Residual Skip Connection}\centering & 2\centering & 3x3\centering  & 128 
    \\
    {03/06/10}\centering & 2\centering & 1x1\centering  & 128 
    \\
    \midrule
    {Residual Skip Connection}\centering & 1\centering & 3x3\centering  & 256 
    \\
    {04/07/11}\centering & 1\centering & 1x1\centering  & 256
    \\
    \bottomrule
    \end{tabular}%
    }
\end{table}%

\begin{table}[!ht]
  \centering
  \caption{ConnectedUNets\Plus\Plus \space architecture details in terms of encoder-decoder and ASPP block. Dilation rate of convolution layers in ASPP block is shown in corresponding braces. ASPP block details also apply to ConnectedUNets\Plus. \centering}
  \renewcommand{\arraystretch}{0.68}
  \scalebox{1}
  {
  \label{table-encoder}
  
    \begin{tabular}{*{6}{c} >{\collectcell\pa}c<{\endcollectcell}}
    \toprule
    \textbf{Block} & \textbf{Layer} & \textbf{Filters} & \textbf{Block} & \textbf{Layer} & \textbf{Filters} \\
    \midrule
    & 3x3 Conv. & 8 & & 3x3 Conv. & 17\\
         Encoder (01/05) & 3x3 Conv. & 17 & Encoder (02/06) & 3x3 Conv. & 35  \\
         Decoder (04/08) & 3x3 Conv. & 26 & Decoder (03/07) & 3x3 Conv. & 53 \\
          & 1x1 Conv. & 51 & &  1x1 Conv. & 105 \\
          
    \midrule
    & 3x3 Conv. & 35 & & 3x3 Conv. & 71  \\
         Encoder (03/07) & 3x3 Conv. & 71 & Encoder (04/08) & 3x3 Conv. & 142 \\
         Decoder (02/06) & 3x3 Conv. & 106 & Decoder (01/05) & 3x3 Conv. & 213  \\
          & 1x1 Conv. & 212 & & 1x1 Conv. & 426 \\
    
    \midrule
    \multirow{5}{*} & 3x3 Conv. & 512 & & 3x3 Conv.  & 32 \\
         ASPP Block\centering & 3x3 Conv.(6) & 512 & ASPP Block\centering & 3x3 Conv.(6)  & 32 \\
         (Bottleneck)\centering & 3x3 Conv.(8) & 512 & (Output Layer)\centering\centering & 3x3 Conv.(8) & 32\\
          & 3x3 Conv.(12) & 512 & & 3x3 Conv.(12) & 32 \\
          & 1x1 Conv. & 512 & & 1x1 Conv. & 32 \\
    
    \bottomrule
    \end{tabular}%
  }
\end{table}%

As mentioned by Ibtehaz et al.~\cite{ibtehaz2020multiresunet} by adding these residual path connections, the proposed architectures are more immune to perturbations, and outliers. They also help to obtain better results in less time and fewer epochs. Additionally, we have used Atrous Spatial Pyramid Pooling (ASPP) blocks to preserve the same bottleneck structure for both architectures as Connected-UNets.
The architectural details of the encoder, decoder and ASPP blocks are described in Table~\ref{table-encoder}.

On the encoder side, each encoder block's output is subjected to a maximum pooling operation before the features are forwarded to the next encoder, and the output of the last encoder passes through an ASPP block before being fetched to the first decoder. Each decoder block is made up of a 2x2 transposed convolution unit that up-samples the preceding block's features before concatenating them with the encoder features received by the residual skip connection; and these features are then fetched to the decoder above. A second U-Net is connected via a new set of residual skip connections, which are utilized to transfer information from the previous U-Net. 
The output of the final decoder block of the first U-Net is fed into a 3x3 convolution layer before it gets concatenated with itself again, followed by an activation ReLU and a BN layer. 
This acts as the first encoder block's input to the second U-Net. The output of the max pooling operations of each of the three encoder blocks is fed into a 3×3 convolution layer and then concatenated with the output of the preceding decoder block of the first U-Net. 

The ASPP block receives the output features of the second U-Net's last encoder block; the remaining blocks are the same as discussed in the first U-Net. Finally, the predicted segmentation mask is generated by passing the output of the last decoder to another ASPP block followed by a 1x1 convolution layer and a sigmoid activation layer. 
In contrast to the work presented in the Connected-UNets paper~\cite{baccouche2021connected}, our work considers a full-fledged mammographic image as input instead of only a ROI since it typically fails to detect micro masses present in the image. 
The ROI extracted segmentation does not help in the real life scenario because the detection and localization of the mask has to be done using a different neural network or manually.

\subsection{Dataset Preprocessing \& Experimental Setup}
We evaluated the proposed architecture on two publicly available datasets, CBIS-DDSM~\cite{lee2017curated} and INbreast~\cite{moreira2012inbreast}. CBIS-DDSM contains 2478 mammography images from 1249 women and included both craniocaudal (CC) and mediolateral oblique (MLO) views for most of the exams of which 1231 cases contain single or multiple breast masses. 
This dataset includes real-world mammograms with background artifacts, poor contrast, corners, borders, and different orientations. To address these issues, we have applied several preprocessing steps which include border cropping to tackle the white border or corner problem, normalizing pixel values to the interval 0 to 1, eliminating background artifacts, and finally, applying CLAHE for contrast enhancement. This improves the mammogram's fine details, textures, and features that would otherwise be challenging for the model to learn. To preprocess the ground truth masks, we discarded the same amount of the borders to get rid of any artifacts and used appropriate padding to make the masks square. In the case of the CBIS-DDSM dataset, after preprocessing and fusing multiple masks of the same image, we split the 1231 images in the training set using an 85:15 ratio for training and validation (i.e., 1046 and 185 images, respectively). The test dataset had 359 images.

INbreast dataset was built with full-field digital mammograms and has a total of 115 cases which include both masses and calcifications. 
In total, the dataset has only 107 images of breasts with masses. We have used 69 images for training, 17 images for validation, and 21 images for testing.
For preprocessing, we have solely applied CLAHE both on the mammograms and the ROIs.

In all of the experiments, adam~\cite{kingma2014adam} optimizer is used with an initial learning rate of 0.0001. 
Batch size of 16 is used during training and testing. We have experimented with input size of 224x224 and 256x256. The best score for all the architectures was obtained with the 224x224 input size, which is reported in the results section. Most of the papers mentioned in the related work section use a mixture of Dice and IOU loss. The primary motivation for directly using this loss is to maximize those metrics. However, this gives no information about convergence. So, to remove confusion regarding the convergence, the loss function used here is binary crossentropy which is the standard loss function for image segmentation task.



\subsection{Evaluation Metrics}
In semantic segmentation, the region of interest typically occupies a small area of the entire image. Therefore, metrics like precision and recall are inadequate and often lead to a false sense of superiority, inflated by the perfection of detecting the background. To evaluate our approach, we have considered four metrics in our experiments: Dice score (F1 score), Jaccard Index (IoU Score), accuracy and Hausdorff distance (H). 
Even though the Dice score (Eq. \ref{eq-dice}) and Jaccard index (Eq. \ref{eq-iou}) are two widely used metrics for semantic segmentation, they are biased towards large masses. The Hausdorff distance (Eq. \ref{eq-haus}) is an unbiased metric that treats all objects equally independently of their size. It measures the maximum deviation along the boundary between the ground truth and predictions. 
\begin{equation}\label{eq-dice}Dice\: score(A,B) =\frac{2\times Area\: of\: Intersection(A,B)}{Area\: of (A) +Area\: of (B)} = \frac{2\times(A\cap B)}{A+B}\end{equation}

\begin{equation}\label{eq-iou}IoU\: score(A,B) = \frac{Area\: of\: Intersection(A,B)}{Area\: of\: Union(A,B)}=\frac{A\cap B}{A\cup B}\end{equation}

\begin{equation}\label{eq-haus}H = max(h(GT, pred),h(pred,GT))\end{equation}

\section{Experimental Results \& Discussion}
To assess the performance, all models have been run for 300 and 400 epochs for the CBIS-DDSM and INbreast datasets, respectively with early stopping and ReduceLROnPlateau callback function. 
Table~\ref{table-cbis-dice} shows the comparison of the proposed architectures' results with some state-of-the-art methods on the CBIS-DDSM dataset. As it can be observed, ConnectedUNets\Plus\Plus \space consistently outperforms the baseline model Connected-UNets and Connected-ResUNets along with other models used for mass segmentation.
ConncetedUNets\Plus \space performed better than Connected-UNets on the test dataset even though the only difference between them is the residual skip connections.

As seen in Table~\ref{table-cbis-dice}, the number of parameters of our proposed methods is higher than the number of parameters of the baseline architectures since we used residual skip connections between encoders and decoders as well as between the two U-Nets. 
Furthermore, all of the architectures performed better on the training and validation sets, but performed poorly on the test set. In particular, the architectures fail to detect any mass from some of the images in the test set, thus giving an output of no ROI and reducing the metric values. We speculate that this is due to the inferior and substandard quality of the scanned images. In Table~\ref{table-cbis-asmavsours}, we compared our proposed architecture with the baseline model by using individual cases from the CBIS-DDSM test dataset considering various thresholds for different metrics. As shown, the proposed architecture was able to predict more cases for each threshold with respect to Connected-ResUNets; however, the average score was better for the baseline architecture. Here we argue that 
ConnectedUNets\Plus\Plus \space has been able to correctly segment smaller masses more accurately than Connected-ResUNets.

\begin{table}[!ht]
  \centering
  \newcommand{\hlc}{\textbf}
  \caption{Comparison of the proposed architectures and state-of-the-art methods on the CBIS-DDSM dataset. Here, DS: Dice score, JI: Jaccard index, Acc.: Accuracy, Param.: No of parameters(in million).\centering}
  \label{table-cbis-dice}
  \scalebox{0.9}
  {
  \begin{tabular}{*{11}{c} >{\collectcell\pa}c<{\endcollectcell}}
    \toprule
    &
    &
    \multicolumn{3}{c}{Training} &
    \multicolumn{3}{c}{Validation} &
    \multicolumn{3}{c}{Test} 
    \\
    \cmidrule(lr){3-5} 
    \cmidrule(lr){6-8}
    \cmidrule(lr){9-11}
    Model Name & 
    Param. & 
    DS & JI & Acc. &
    DS & JI & Acc. &
    DS & JI & Acc.\\
    \midrule
    U-Net\centering & 
        7.8\centering & 
        0.73 & 0.57 & 99.85 & 
        0.73 & 0.58 & 99.87 & 
        0.41 & 0.27 & 99.69 \\
    
    \addlinespace
    MultiResUNet\centering & 
        7.3\centering & 
        0.74 & 0.59 & 99.89 & 
        0.76 & 0.61 & 99.88 & 
        0.40 & 0.26 & \textbf{99.7} \\
        
    \addlinespace
    AUNet\centering & 
        11.01\centering & 
        \textbf{0.89} & \textbf{0.81} & \textbf{99.94} & 
        \textbf{0.90} & \textbf{0.82} & \textbf{99.94} & 
        0.46 & 0.31 & 99.69 \\
        
    \addlinespace
    Connected-UNets\centering & 
        20.1\centering & 
        0.81 & 0.68 & 99.90 & 
        0.81 & 0.68 & 99.91 & 
        0.40 & 0.27 & 99.69 \\
        
    \addlinespace
    ConnectedUNets\Plus\space(ours)\centering & 
        23.5\centering & 
        0.78 & 0.64 & 99.88 & 
        0.77 & 0.63 & 99.89 & 
        0.44 & 0.30 & \textbf{99.7} \\
        
    \addlinespace
    Connected-ResUNets\centering & 
        20.7\centering & 
        0.84 & 0.73 & 99.91 & 
        0.84 & 0.72 & 99.92 & 
        0.47 & 0.32 & 99.69 \\
    
    \addlinespace
    ConnectedUNets\Plus\Plus\space(ours)\centering & 
        28.15\centering & 
        0.88 & 0.79 & \textbf{99.94} & 
        0.88 & 0.79 & \textbf{99.94} & 
        \textbf{0.48} & \textbf{0.33} & \textbf{99.7} \\
        
    \bottomrule
  \end{tabular}
  }
\end{table}

\begin{table}[!ht]
  \centering
  \newcommand{\hlc}{\textbf}
  \centering
  \caption{Comparison of prediction on the CBIS-DDSM test dataset and correctly predicted number of cases over multiple thresholds of Dice score (DS), Jaccard index (JI) and Hausdorff distance (HD) metric between the baseline architecture and our proposed architecture. Here, NC and AS represent the number of cases and average score, respectively.\centering} 
  \label{table-cbis-asmavsours}
  \begin{tabular}{p{3.3cm}*{10}{p{.75cm}}>{\collectcell\pa}c<{\endcollectcell} }
    \toprule
    &
    
    \multicolumn{2}{p{1.50cm}}{DS $\geq$0.45\centering} &
    \multicolumn{2}{p{1.50cm}}{DS $\geq$0.65\centering} &
    \multicolumn{2}{p{1.50cm}}{JI $\geq$0.35\centering} &
    \multicolumn{2}{p{1.50cm}}{JI $\geq$0.55\centering} &
    \multicolumn{2}{p{1.50cm}}{HD $\leq$2.75\centering} 
    \\
    \cmidrule(lr){2-3}\cmidrule(lr){4-5}\cmidrule(lr){6-7}\cmidrule(lr){8-9}\cmidrule(lr){10-11} 
    
    Architecture Name\centering & 
    NC\centering & AS\centering & 
    NC\centering & AS\centering &
    NC\centering & AS\centering &
    NC\centering & AS\centering &
    NC\centering & {AS\centering}\\
    
    \midrule
    
    Connected-ResUNets\centering & 
        160\centering & \textbf{0.78}\centering & 
        134\centering & \textbf{0.82}\centering & 
        154\centering & \textbf{0.66}\centering & 
        119\centering & \textbf{0.73}\centering & 
        174\centering & \textbf{1.85} \\
    
    \addlinespace
    {ConnectedUNets\Plus\Plus}\space(ours)\centering & 
        \textbf{181}\centering & 0.76\centering & 
        \textbf{143}\centering & 0.81\centering	& 
        \textbf{172}\centering & 0.64\centering & 
        \textbf{123}\centering & 0.71\centering & 
        \textbf{191}\centering & 1.92 \\
        
    \bottomrule
  \end{tabular}
\end{table}

\begin{table}[!ht]
  \centering

  \caption{Comparison of the proposed architectures and state-of-the-art methods on INbreast dataset.\centering}
  \label{table05}
  \scalebox{0.9}
  {
  \begin{tabular}{*{11}{c} >{\collectcell\pa}c<{\endcollectcell}}
    \toprule
    &
    &
    \multicolumn{3}{c}{Training} &
    \multicolumn{3}{c}{Validation} &
    \multicolumn{3}{c}{Test} 
    \\
    \cmidrule(lr){3-5}\cmidrule(lr){6-8}\cmidrule(lr){9-11}
    
    Model Name\centering & 
    Param. & 
    DS & JI & Acc. &
    DS & JI & Acc. &
    DS & JI & Acc.\\
    \midrule
    U-Net\centering & 
        7.8\centering & 
        0.87 & 0.77 & 99.91 & 
        0.91 & 0.83 & 99.91 & 
        0.91 & 0.84 & 99.91 \\
    
       
    \addlinespace
    AUNet\centering & 
        11.01\centering & 
        0.94 & 0.89 & 99.94 & 
        0.96 & 0.90 & 99.93 & 
        0.94 & 0.89 & 99.94 \\
        
    \addlinespace
    Connected-UNets\centering & 
        22.4 \centering & 
        0.94 & 0.89 & \textbf{99.99} & 
        0.97 & 0.95 & 99.95 & 
        0.97 & 0.94 & \textbf{99.99} \\
        
    \addlinespace
    ConnectedUNets\Plus \space(ours)\centering & 
        23.5\centering & 
        \textbf{0.97} & 0.94 & 99.97 & 
        \textbf{0.98} & 0.96 & 99.97 & 
        0.98 & 0.95 & 99.97 \\
        
    \addlinespace
    Connected-ResUNets\centering & 
        20.7\centering & 
        0.94 & 0.88 & \textbf{99.99} & 
        0.96 & 0.93 & \textbf{99.98} & 
        0.97 & 0.94 & \textbf{99.99} \\
    
    \addlinespace
    ConnectedUNets\Plus\Plus \space(ours)\centering & 
        28.15\centering & 
        \textbf{0.97} & \textbf{0.95} & \textbf{99.99} & 
        \textbf{0.98} & \textbf{0.97} & \textbf{99.98} & 
        \textbf{0.99} & \textbf{0.97} & \textbf{99.99} \\
        
    \bottomrule
  \end{tabular}
  }
\end{table}

\begin{figure}[!t]
\centering
\includegraphics[width= 1\textwidth]{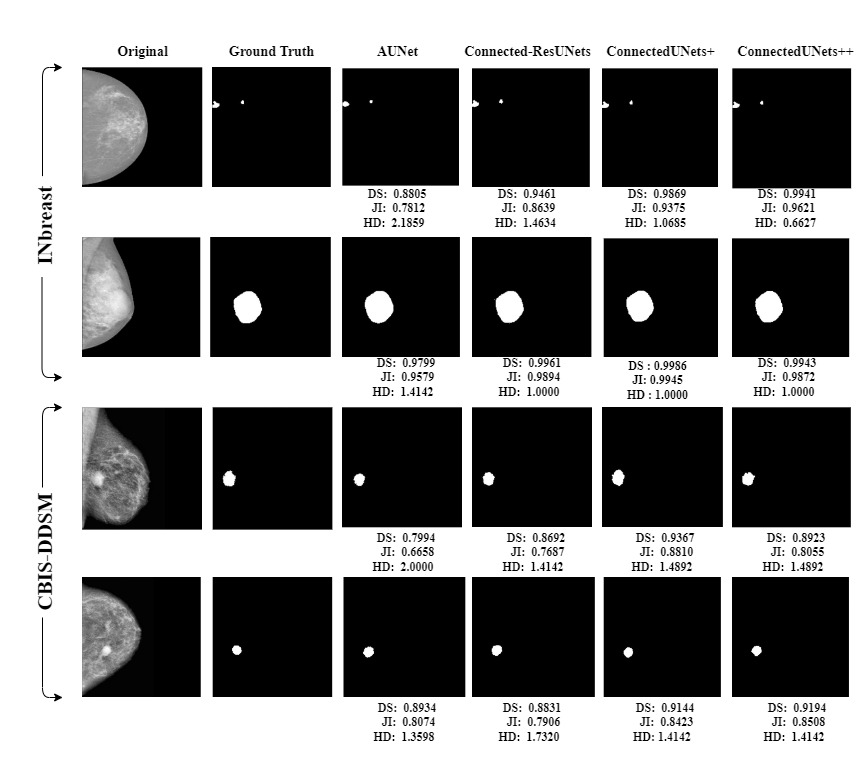}
\caption{Segmentation results of different networks. From left to right, the columns correspond to the input images, the ground truth labels, the segmentation results of AUNet, Connected-ResUNets, ConnectedUNets\Plus\space and ConnectedUNets\Plus\Plus, respectively on the INbreast and CBIS-DDSM dataset.\centering} \label{fig3}

\end{figure}

In Table~\ref{table05}, we also compared the proposed architectures against the baseline and other models on the INbreast dataset. As it can be observed, our model performs noticeably better on the INbreast dataset. We hypothesize that this is due to the two datasets' disparate image quality. The mammograms in the CBIS-DDSM dataset have been scanned, hence the images are of poor quality. The images in the INbreast dataset, however, have been digitally enhanced, and their quality is outstanding. Due to page restrictions, we had to omit comparison of correctly predicted cases like Table~\ref{table-cbis-asmavsours} for the INbreast dataset.

The introduction of residual skip connections between the encoder and decoder has had a major impact on the segmentation task. The obscure and vague boundaries which other architectures fail to correctly segment (either under-segment or over-segment), are properly segmented by ConnectedUNets\Plus \space and ConnectedUNets\Plus\Plus. Additionally, in some complex cases, because of the quality and nature of the ROIs, it becomes challenging to segment homogeneous ROIs. Even in those cases, 
ConnectedUNets\Plus\Plus \space exceeds other architectures in terms of Dice score, Jaccard index, and Hausdorff distance. As Hausdorff distance is highly recommended for cases with complex boundaries, our results show that the proposed architecture can predict mass boundaries more accurately. Since the boundary shape of a mass is a strong indicator of benign and malignant cases~\cite{mahmood2022breast}, the proposed architecture is more suited for mass prediction and segmentation in real-world scenarios.

Segmentation examples both for the INbreast and CBIS-DDSM datasets are shown in Figure~\ref{fig3}. We have compared the segmentation result of ConnectedUNets\Plus\space and ConnectedNets\Plus\Plus\space with two best-performed architectures: AUNet and Connected-ResUNets. In all the cases, the proposed architectures outperformed all other architectures and achieved almost perfect scoring regardless of the ROI's shape. 
Figure~\ref{fig3} showcases how well the models worked not just for the larger ROI (Case 2) but also for those with smaller ROI (Case 1) on INbreast dataset. The models yield similar results for CBIS-DDSM dataset as well.

\section{Conclusion}
We started this work by thoroughly examining the Connected-UNets architecture in order to identify potential areas for improvement. In this context, we identified certain inconsistencies between the encoder's and decoder's features. Inspired by MultiResUNet, we added  some additional processing between them to make them more homogeneous. 
Furthermore, to give Connected-UNets the capacity to perform multi-resolutional analysis, we introduced residual blocks into the encoder and decoder architecture which resulted in ConnectedUNets\Plus\Plus, a novel architecture that incorporates these changes. 
Unlike architectures previously discussed in the literature, we have used all of the available images for training and testing. 
For future work, we plan to conduct additional experiments to determine the ideal choice of nodes, layers, and hyperparameters. Furthermore, we would also like to assess the efficacy of our model on datasets of medical images from various modalities. 

\subsubsection{Acknowledgements} 
Portions of this material is based upon work supported by the Office of the Under Secretary of Defense for Research and Engineering under award number FA9550-21-1-0207.

%
%
%
\bibliographystyle{ieeetr}
\bibliography{bibliography.bib}

\end{document}